# Electromagnetic approach to ultrathin solar cell efficiencies

A. Niv[1], M. Gharghi[1], Z.R. Abrams[1], C. Gladden[1], and X. Zhang[1,2]

[1]NSF Nanoscale Science and Engineering Center (NSEC), 3112 Etcheverry Hall, University of California, Berkeley, California 94720, USA
[2]Materials Science Division, Lawrence Berkeley National Laboratory, 1 Cyclotron Road, Berkeley, California 94720, USA

Current methods for evaluating solar cell efficiencies cannot be applied to extremely thin cells where phenomena from the realm of near field optics prevail. We overcome this problem by offering a rigorous electromagnetic calculation of solar cell efficiencies based on the fluctuation dissipation theorem. Our approach is demonstrated by calculating the efficiency of a GaAs solar cell with an Au back reflector for thicknesses well below the typical wavelength of the solar flux. It is shown that near field optics affect the performance of low dimensional solar cells.

The demand for cheaper and more efficient solar cells has prompted the consideration of thinner devices [1,2]. The principle concern of thinner cells lies in the reduction of their absorption, which directly affects the electric current extracted from the cell. In order to overcome this problem researchers have turned to the field of near field optics, which has been successful in guiding and localizing light well below the dimensions of their free space wavelengths. Examples in the realm of solar cells include the theoretical prediction of high absorbance in systems as thin as 5 *nm* [3] and operating solar cells with a thickness of only 160 *nm* [4]. In particular, near field effects incorporating plasmonic enhancement have been suggested as means to locally concentrate light [5-7], or for optimal light scattering designs [8-11].

While near field optics techniques primarily target enhanced absorption in order to increase the photo-current, solar cell design must be optimized for output *power* – the product of current *and* voltage. This voltage is usually analyzed using the thermodynamics of the cell, for which established theories exist for cells that are much thicker than the typical wavelength of the solar spectrum [12-17]. However, being based on ray optics these theories fail for ultrathin cells, where near field optics prevails. Here, we present a rigorous analysis of the solar cell performance at dimensions well below optical wavelengths. We use basic Electromagnetism (EM) in conjunction with the Fluctuation Dissipation Theorem (FDT) [18] to quantify the rate of emission from the cell. This enables the calculation of the voltage, current, and therefore the efficiency of an ultra thin cell. We demonstrate our approach by calculating the open circuit voltage and efficiency for a simple GaAs slab cell with an Au back reflector, with dimensions as thin as few nanometers.

The ideal efficiency of a solar cell is predominantly characterized by the detailed balance between absorbed and emitted photons [12]. The absorption rate for a semiconductor with a bandgap of $E_G=\hbar\omega_G$ is simply given by:

$$R_{IN} = \frac{\Omega_S}{4\pi^3 c^2} \int_{\omega_G}^{\infty} \frac{\sigma(\omega)\omega^2 d\omega}{\exp(\hbar\omega/k_B T_S)-1} \tag{1}$$

where $\Omega_S = 5.86 \times 10^{-5}$ $sr$ is the solid angle subtended by the sun, $c$ is the speed of light, $\omega$ is the angular frequency, $\hbar$ is the reduced Plank's constant, $k_B$ is Boltzmann's constant, $T_S = 5800K$ is the temperature of the sun, and $\sigma(\omega)$ is the power absorbance of the cell. To calculate the emission rate we first assume that rapid diffusion of carriers in the semiconductor makes for uniform carrier density [12]. Under this condition spatial and spectral variations in absorption do not affect the emission, and the two can be treated independently. This assumption is justified as long as the optical wavelength at hand is smaller than the carrier diffusion length, which is typical in most solar cell materials. The emission rate is found from the power flux emanating from the surface of the cell:

$$R_{OUT} = \int_A da \int_0^{\infty} \frac{d\omega}{\hbar\omega} \langle \mathbf{S}(\mathbf{r},\omega) \rangle \cdot \hat{\mathbf{n}} \tag{2}$$

where $\hat{\mathbf{n}}$ is the surface normal, and the ensemble average of the Poynting vector, $\langle \mathbf{S} \rangle$, is given in the Green function formalism by:

$$\langle S_i(\mathbf{r}_1,\omega) \rangle = \int_V dr_2^3 \varepsilon_{ijk} G_{jm}^E(\mathbf{r}_1-\mathbf{r}_2;\omega) P_{ml}(\mathbf{r}_1,\mathbf{r}_2;\omega) G_{lk}^{H*}(\mathbf{r}_1-\mathbf{r}_2;\omega) + c.c. \tag{3}$$

where $G_{jm}^E$ and $G_{lk}^{H*}$ are the electric and magnetic Green dyads of the system, $\varepsilon_{ijk}$ is the Levi-Civita tensor, *c.c.* denotes complex conjugate, and summation over repeated indices is assumed. From FDT [18] the stochastic source for EM radiation, $P_{ml}(\mathbf{r}_1,\mathbf{r}_2;\omega)$, is given by:

$$P_{ml}(\mathbf{r}_1,\mathbf{r}_2;\omega) = (\hbar\omega/2)\coth(\hbar\omega/2k_B T_o)\varphi_{ml}(\mathbf{r}_1,\mathbf{r}_2;\omega) \tag{4}$$

with $\varphi_{ml}$ as the dissipation function.

In order to identify the correct form for $\varphi$, we note that the spontaneous emission rate, which dominates radiative recombination at optical frequencies, is given for a semiconducting material at temperature $T_o$ by [19]:

$$\dot{N}_{SPONT} = \left(\frac{\varepsilon'\omega^2}{\pi^2 c^2}\right)\left[\exp\left(\frac{\hbar\omega - \mu}{k_B T_o}\right) - 1\right]^{-1} \alpha(\omega) \quad (5)$$

Here, $\varepsilon=\varepsilon'+i\varepsilon''$ is the material permittivity, $\mu$ is the chemical potential, and $\alpha$ is the dissipation of the material at hand. For low loss materials ($\varepsilon'>\varepsilon''$) $\alpha$ is given by:

$$\alpha(\omega) = \frac{\omega}{c}\frac{\varepsilon''}{\sqrt{\varepsilon'}} \quad (6)$$

Inserting Eq. (6) into Eq. (5) and comparing to the known expression for the emission rate from a dipole in free space (Larmor formula), the appropriate form for the response function is:

$$\varphi_{ml}(\mathbf{r}_1, \mathbf{r}_2; \omega) = \frac{\varepsilon''(\mathbf{r}_1, \omega)}{\omega} \frac{1}{\exp((\hbar\omega-\mu)/k_B T_o) - 1} \delta(\mathbf{r}_1 - \mathbf{r}_2)\delta_{ml} \quad (7)$$

Were $\delta(\cdot)$ and $\delta_{ij}$ are the Dirac and Kronecker delta functions, respectively. Eq. (7) ensures that fluctuations of charge-polarization emerge as the stochastic EM sources in the right hand side of Eq. (4).

Equations (2)-(4) along with Eq. (7) establish a rigorous EM algorithm for the determination of the emission rate from a solar cell. Depending on the cell structure, the modified density of optical states in the near field regime may increase or decrease the emission from EM sources. Nevertheless, it is the out-coupling of the power following Eq. (2) that determines the photon flux emerging from the device. As such, the presented approach provides an invaluable methodology to assess and optimize the performance of ultrathin solar cells.

Solar cell characteristics can be calculated based on the absorption and emission rates in Eqs. (1) and (2): The current is given by: $I = q[R_{IN} - R_{OUT}(V)]$, with the voltage proportional to the chemical potential according to $qV=\mu$, with $q$ as the electronic charge. The maximal efficiency is obtained by maximizing the $I \times V$ product, divided by the input power [17].

In order to demonstrate our approach, we analyze the simplified cell shown in Fig. (1), whose structure was chosen to highlight optical near field effects while assuming other mechanisms such as charge separation and collection to be trivial. Although not a practical device by itself, simplicity and the high degree of symmetry permit a near analytical evaluation of the emission rate. Since the solar spectrum can be safely regarded as emanating from a plane wave source no near field effects are present for the absorption ,and for the absorption alone, and $\sigma(\omega)$ is calculated from closed form formulas [20,21].

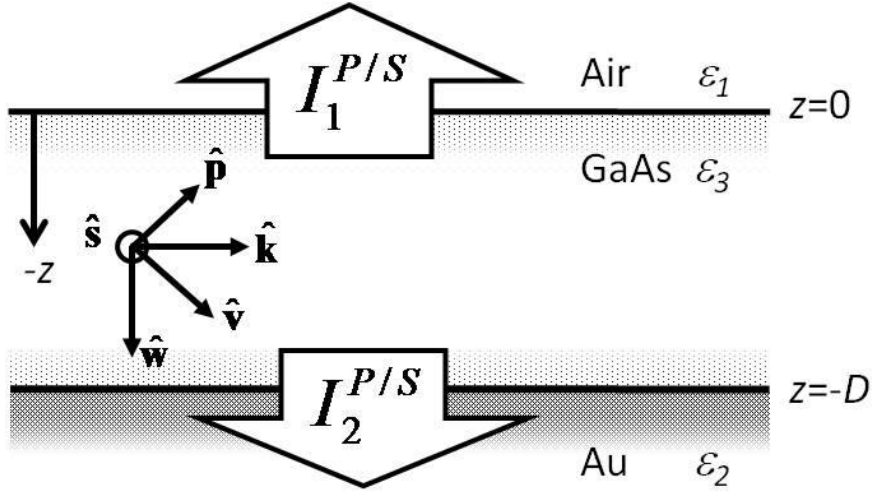

**Figure 1. Schematic layout of the solar cell model analyzed in the text. A slab of GaAs lies between a metallic Au layer, and air. Block arrows represent the four possible emission channels: GaAs to air - $I_1^{P/S}$, and GaAs to Au substrate - $I_2^{P/S}$, for *TE* (*S*) and *TM* (*P*) polarizations. Notations in the figure are as used and defined in the text.**

The emission rate is calculated using Sipe's expression for the Green Dyads of this structure [21] and after some manipulation assumes the form:

$$R_{OUT} = \frac{1}{2\pi^2 c}\exp\left(\frac{\mu}{k_B T_o}\right)\int_0^\infty d\omega\left[\omega\exp\left(\frac{-\hbar\omega}{k_B T_o}\right)\int_0^\infty kdk\left(I_1^S + I_1^P + I_2^S + I_2^P\right)\right] \quad (10)$$

here *k* is a strictly real wavenumber tangent to the interfaces, and the notation is used as in Fig. (1). For simplicity, bulk permitivities are adopted from Refs. [22] and [23]. We have used the standard approximation that: $\hbar\omega$-$\mu$>>$k_B T_o$ and thus $coth(\hbar\omega/2k_B T_o)\approx 1$ as well as dropping the -1 in the denominator of Eq. (7). The four emission channels of Eq. (10) are related to the two interfaces of the GaAs slab, and the two polarizations. The coresponding integrands are given by:

$$I_1^S = \frac{|\tau_1^S|^2}{|w_3|^2}\left(\frac{\varepsilon_1' w_1'}{|v_1|}\right)\left[\frac{w_3'}{2}\left(1-e^{-2w_3''D}\right)\left(1+|r_{32}^s|^2 e^{-2w_3''D}\right) + 2w_3'' e^{-2w_3''D}\,\mathrm{Im}\left\{r_{32}^s\left(e^{2iw_3'D}-1\right)\right\}\right] \quad (11a)$$

$$I_1^P = \frac{|\tau_1^P|^2}{|w_3|^2}\left(\frac{\varepsilon_1' w_1'}{|v_1|}\right)\left[\frac{w_3'}{2}\left(1-e^{-2w_3''D}\right)\left(1+|r_{32}^P|^2 e^{-2w_3''D}\right) + 2w_3''\frac{|w_3|^2 - k^2}{|v_3|^2} e^{-2w_3''D}\,\mathrm{Im}\left\{r_{32}^P\left(e^{2iw_3'D)}-1\right)\right\}\right] \quad (11b)$$

$$I_2^S = \frac{|\tau_2^S|^2}{|w_3|^2}\left(\frac{\varepsilon_2' w_2'}{|v_2|}\right)\left[\frac{w_3'}{2}\left(1-e^{-2w_3''D}\right)\left(1+|r_{31}^s|^2 e^{-2w_3''D}\right) + 2w_3'' e^{-2w_3''D}\,\mathrm{Im}\left\{r_{31}^s\left(1-e^{-2iw_3'D}\right)\right\}\right] \quad (11c)$$

$$I_2^P = \frac{|\tau_2^P|^2}{|w_3|^2}\left(\frac{\varepsilon_2' w_2'}{|v_2|}\right)\left[\frac{w_3'}{2}\left(1-e^{-2w_3''D}\right)\left(1+|r_{31}^P|^2 e^{-2w_3''D}\right)+2w_3''\frac{|w_3|^2-k^2}{|v_3|^2}e^{-2w_3''D}\operatorname{Im}\{r_{31}^P(1-e^{-2iw_3'D})\}\right] \quad (11d)$$

Here $\mathbf{v}_i = \sqrt{\varepsilon_i}(\omega/c)\hat{\mathbf{v}}_i = k\hat{\mathbf{k}} + w_i\hat{\mathbf{z}}$ is the wave vector with the longitudinal wave number $w_i = \sqrt{\varepsilon_i(\omega/c)^2 - k^2}$ ($i=1,2,3$ for Air, Au, GaAs respectively), $r_{ji}^{P/S}$, $t_{ji}^{P/S}$ denote Fresnel's reflection and transmission coefficients from material $i$ into material $j$ for optical polarizations $p$ or $s$, respectively, and $D$ is the (varying) thickness of the GaAs slab. Prime and double-prime symbols denote real and imaginary parts of a complex quantity. The first term in the square brackets of Eqs. (11a) through (11d) represents a direct contribution from a specific emission source plane within the GaAs and its reflection from the opposite interface; the second results from interference of the former two. For $p$ polarization this interference term is augmented by an additional term resulting from the relative angle between the EM polarizations of the two waves. The term in parentheses to the left of the integrals is the projection of the Poynting vector onto the surface normal, and transmission and multiple reflections are accounted for by:

$$\tau_{1/2}^{P/S} = \frac{t_{31 or 32}^{P/S}}{1 - r_{31}^{P/S} r_{32}^{P/S} \exp(-w_3''D)} \quad (12)$$

It can be shown that for $D \to \infty$, $t_{31}^{P/S} \to 1$, and by neglecting losses, Eq. (10) reduces to Planck's well-known formula for black body radiation from a semiconducting material. An important aspect of our formalism is that the existence of a band gap enters generically by Eqs. (11) rather than being artificially imposed as in Eq. (1).

Figure 2(a) depicts the open circuit voltage, $V_{OC}$, versus cell thickness for the device in Fig. (1) and for cell temperature of $T_o=300$ K. This voltage can be calculated by extracting the chemical potential from Eq. (10) at open circuit equilibrium where $I=0$ and can be approximated as [13,15]:

$$V_{OC} = \frac{\mu_{OC}}{q} \cong \frac{k_B T_o}{q}\ln\left(\frac{R_{IN}}{R_{OUT}(\mu=0)}\right) \quad (13)$$

The quasi-periodic oscillations in $V_{OC}$ result from interference patterns in both $R_{IN}$ and $R_{OUT}$ and tend to relax as the cell's thickness increases. For a thick cell, our analysis yields $V_{OC}=$ 1.14V (inset in Fig. 2(a)), in agreement with predictions of previous models [12]. The efficiency shown in Fig. 2(b) is calculated by: $\eta = FF \times V_{OC} \times I_{SC}/P_{IN}$, where $P_{IN}$ is the total available solar power, and the fill factor $FF$ is approximated using the empirical relation from Ref. [17]. We obtain a saturated efficiency of 18.87% for thick cells (data not shown), while for thin cells the efficiency drops as a result of reduced absorption in the GaAs slab.

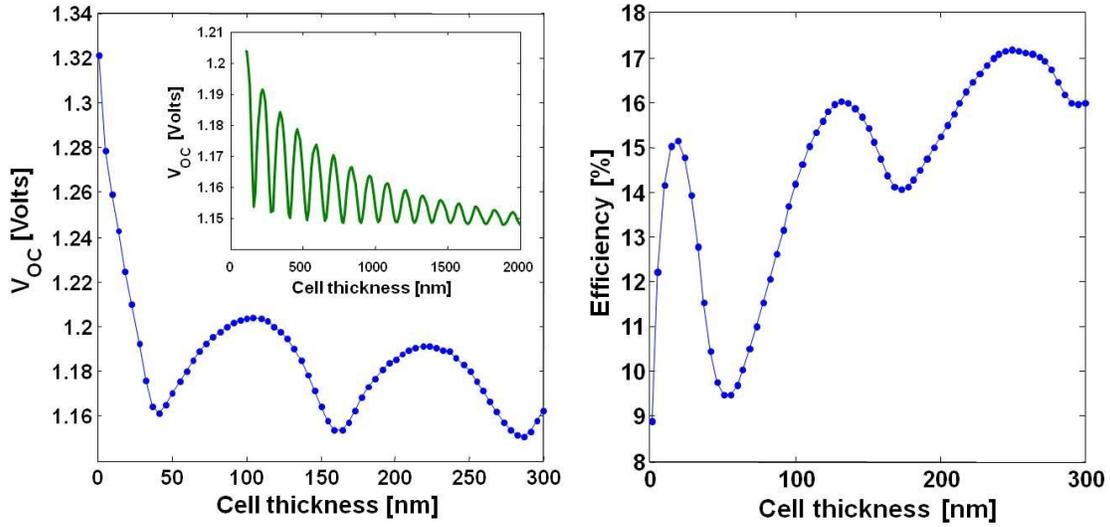

**Figure 2:** Open circuit voltage (a) and overall efficiency (b) versus solar cell thickness, using the schematic of Fig. 1. Oscillations can be seen in both panels due to the etalon-generated interference patterns within the GaAs slab due to numerous frequencies. Inset in (a): Damping of the oscillations for thick cells.

Although the above example does not represent a viable design, as the efficiency drops sharply with cell thickness, it can be modified to indicate the potential of ultrathin cells. For this, let us assume perfect absorption in Eq. (1), i.e. $\sigma(\omega)=1$. We note that without introducing further losses, this offers superior enhancement absorption over any light trapping scheme [16], and thus presents the maximal efficiency. For completeness let us also assume a perfectly matched boundary at the GaAs/Air interface so that no reflections arise, i.e. $r_{31}^{P/S}=0$ and $t_{31}^{P/S}=1$ in Eqs. (11) and (12). The results of these assumptions are plotted for $V_{OC}$ in Fig. 3 as a function of cell thickness.

The evaluation of $V_{OC}$ for an ultrathin cell has never been demonstrated before, and our methodology allows a direct calculation to be made. The increase in $V_{OC}$ as a logarithmic function of the thickness, $D$, has been empirically found to be [2,24]: $dV_{OC}/dD=(k_BT/q)\times(-1/D)$, however, this heuristic formula is only relative to a known value for $V_{OC}$. Our analysis allows a direct calculation of the $V_{OC}$, as well as the overall efficiency (inset, Fig. 3), showing the same logarithmic rise in $V_{OC}$, as calculated using EM, as opposed to using carrier density approximations. Deviation from the pure logarithmic trend results from the wave nature of light and are more pronounced where the optical near fields dominates the emission rate of Eq. (10), i.e. *D<100 nm*.

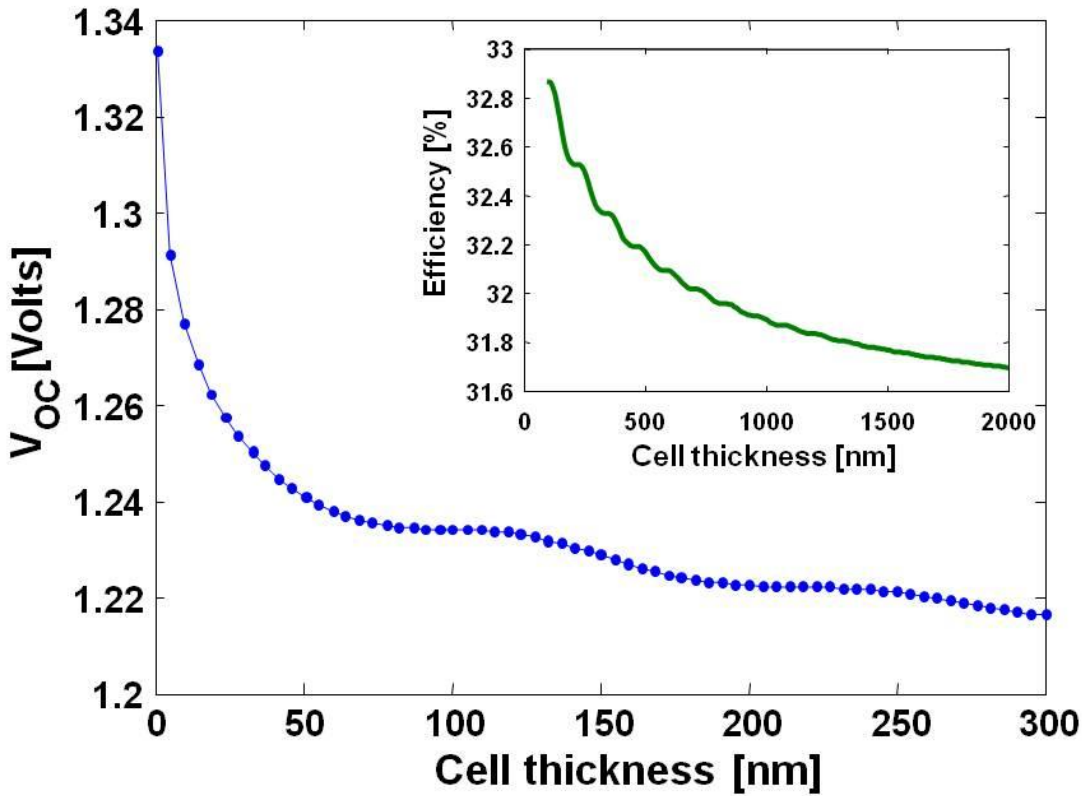

**Figure 3: Open circuit voltage versus solar cell thickness for perfect absorption and a matched GaAs/Air interface. The logarithmic trend that emerges is similar to that derived from charge concentration considerations. Small ripples in the voltage are a result of optical interference effects. Inset: Efficiency for cell thicknesses up to 2 *μm*, demonstrating the saturation value.**

In conclusion we have used a combination of EM and FDT to relate the thermodynamic and the EM aspects of power generation in solar cells operating in the optical near-field regime. This gives an EM framework for evaluating solar cell performance under conditions unattainable with previous approaches. The analysis is not principally limited to semiconductors and can be applied to any system that can be described with the macroscopic Maxwell equations. In addition, since the analysis relies on the EM power emitted from the cell, it is not limited to simple geometries and can be applied to complex structures using commercial EM simulators.

**Acknowledgments**

This work was supported by the U.S. Department of Energy, Basic Energy Sciences Energy Frontier Research Center (DoE-LMI-EFRC) under award DOE DE-AC02-05CH11231. AN and ZRA would like to thank Prof. D. Chandler for his helpful discussion.

**References:**

1. M.A. Green, Physica E **14**, 65 (2002).
2. H.J. Queisser, Physica E **14**, 1 (2002).
3. Z. Yu, A. Raman, and S. Fan, P. Natl. Acad. Sci. USA **107**, 17491 (2010).
4. V.E. Ferry, M.A. Verschuuren, H.B.T. Li, E. Verhagen, R.J. Walters, R.E.I. Schropp, H.A. Atwater, and A. Polman, Opt. Express **18**, A238 (2010).
5. K. R. Catchpole and A. Polman, Opt. Express **16**, 21793 (2008).
6. H. A. Atwater and A. Polman, Nat. Mater. **9**, 205 (2010).
7. S. Maier, *Plasmonics: Fundamentals' and Applications*, (Springer-Verlag, New York 2007).
8. V. E. Ferry, L. A. Sweatlock, D. Pacifici and H. A. Atwater, Nano Lett. **8**, 4391 (2008).
9. R. A. Pala, J. White, E. Barnard, J. Liu and M. L. Brongersma, Adv. Mater. **21**, 3504 (2008).
10. C. Hagglund and B. Kasemo, Opt. Express **17**, 11944 (2009).
11. P. N. Saeta, V. E. Ferry, D. Pacific, J. N. Munday and H. A. Atwater, Opt. Express **17**, 20975 (2009).
12. W. Shockley and H.J. Queisser, J. Appl. Phys. **32**, 510 (1961).
13. R.T. Ross, J. Chem. Phys. **46**, 4590 (1967).
14. W. Ruppel and P. Würfel, IEEE Trans. Electron Devices **27**, 877 (1980).
15. T.T. Tiedje, E. Yablonovitch, G.D. Cody, and B.G. Brooks, IEEE Trans. Electron Devices **31**, 711 (1984).
16. E. Yablonovitch and G. D. Cody, IEEE Trans. Electron Devices **29**, 300 (1982).
17. M. Green, *Solar Cells* (Prentice-Hall Inc., Englewood Cliffs 1998).
18. R. Kubo, M. Toda, and N. Hashitsume, *Statistical physics II* (Springer-Verlag, Berlin 1995), 2[nd] ed.
19. G. Lasher and F. Stern, Phys. Rev. **133**, A553 (1964).
20. M. Born and E. Wolf, *Principles of Optics*, (Cambridge University Press, New York, 1999), 7[th] ed.
21. J.E. Sipe, J. Opt. Soc. Am. B **4**, 481 (1987).
22. P. B. Johnson and R.W. Christie, Phys. Rev. B **6**, 4370 (1972).
23. E.D. Palik, *Handbook of the optical constants of solids* (Academic Press 1998).
24. R. Brendel and H.J. Queisser, Sol. Energ. Mat. Sol. C. **29**, 397 (1993).